\documentclass{iaus}
\usepackage{graphicx}
\usepackage{amsmath}
\usepackage{mathabx} %wasysym} 
\usepackage{sidecap}

\title[Flow of Planets Raises Fall Off] %Explains Distribution% here short title %%
{Flow of Planets Raises Short Period Fall Off}
\author[Stuart F. Taylor]%\& Givennameb Surnameb] %% give here short author list %%
{Stuart F. Taylor$^{1,2}$}
\affiliation{$^1$Job Seeking\\% 
$^2$Participation Worldscope/Global Telescope Science\\
Los Angeles, California, U.S., and Hong Kong, SAR China\\
email: {\tt astrostuart@gmail.com} \\[\affilskip]
}

\pubyear{2012}  % or will it be 2013?
\volume{293}  %% insert here IAU Symposium No.
\pagerange{001--003} %119--126}  Nader, don't the editors have to input this?
\jname{Formation, Detection, and Characterization of Extrasolar Habitable Planets}
\editors{N. Haghighipour, ed.}% A.C. Editor, B.D. Editor \& C.E. Editor, eds.}
\begin{document}

\maketitle

\begin{abstract}

After finding more planets than expected at the shortest period, there
has been an effort to explain their numbers by weak tidal friction. 
However, we find that the strength of tidal dissipation that 
would produce the occurence
distribution found  
from Kepler planet candidates is different for
giant versus medium radii planets. 
This discrepancy can be resolved 
if there is a ``flow'' of the largest planets regularly arriving
such that they go through a ``hot Jupiter'' stage.
We also show a correlation of higher stellar Fe/H with
higher eccentricity of giant planets that may be from
smaller planets having been sent into the star by the migration of
the larger planet. 
This disruption of the orbits of medium and smaller planets 
could account for the lower occurrence of ``hot Neptune'' medium radius planets.
%  which is at http://www.iau.org/static/scientific_meetings/authors/Keywords.txt
\end{abstract}

\firstsection % if your document starts with a section,
\section{Introduction}

The most unexpected planet finding could be 
%is arguably 
that the number of planets
with shorter periods is larger than  expected. Most such close 
planets would been expected to quickly migrate into the star in 
timescales short relative to the lifetime of the stars
unless tidal dissipation is unexpectedly weak, 
leading to work seeking to explain weak dissipation in stars. 
We show how the occurrence distribution of differently-sized planets 
is more consistent with the explanation that these planets have more recently arrived as a flow of inwardly migrating planets, with giant planets more likely to be found while gradually going through stage having the shortest periods. 
This ``flow'' of inwardly migrating planets would be the final stage
%arriving from further out could be supplied by 
of the high eccentricity
migration flow proposed by 
\cite[Socrates et al. (2012; H12 hereafter)]{soc12} %\citep{soc12}.
to explain the short period pileup 
of giant planets.  %(Socrates et al. 2012). 

We have previously shown that the shortest period region of the exoplanet occurrence distribution has a fall-off shaped by inward tidal migration due to stellar tides, that is, tides on the star caused by the planets 
(\cite[Taylor 2012; T12 hereafter]{tay12}). 
%For two refs see hist or pieces
The power index of the fall-off of giant 
(8 to 32 earth radii ``$R_\Earth$'') %short: (8 to 32 $R_\Earth$)
and medium 
(4 to 8 $R_\Earth$)
radii planet candidates found from Kepler data by H12
%\cite[Howard et al. (2012; H12 hereafter)]{how12},
%\citep{how12}.
%(Howard et al. 2012) 
is close to 13/3, which is the power index resulting from tidal migration
(T12). However, there is a discrepancy of the strength of the tidal migration determined using giant and medium planets that is best resolved by the explanation that more giant than medium radii planets are observed passing through these shortest period orbits. 

We also present a correlation between higher eccentricity of planetary orbits with higher Fe/H of host stars, which could be explained by high eccentricity planets being associated with recent episodes of other planets into stars. 
%By the time these planets migrate to become hot Jupiters, 
%the pollution may be mixed into the star. 
The shortest period planet hosting stars may have already convected
the pollution away from the surface.
The clearing of other planets by migrating hot giant planets may result in 
the absence of additional planets in hot Jupiter systems.

%Significant paper to the role of stellar tides causing short period planets
%to migrate into star, by 
%\cite[Jackson et al.(2009)]{jac09}
%\cite[Anders \& Zinner (1993)]{AndersZinner93} and 
%\cite[Ott (1993)]{Ott93}.
%A book reference is \cite[Zinner (2004)]{Zinner04}. 

\begin{figure}[t]
% \vspace*{-2.0 cm}
\begin{center}
 \includegraphics[width=3.4in]{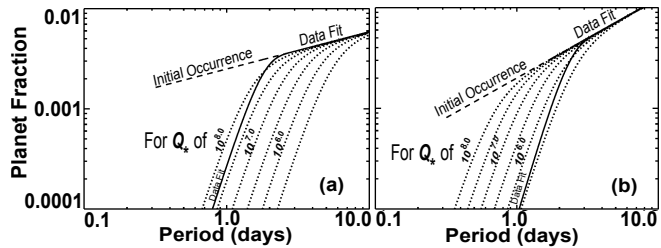}
% Old plotpdistri_lookforQ_Mp159+31Me.eps
% \vspace*{-1.0 cm}
 \caption{Migrated evolution compared to data for masses and radii for
    giant ($100$ to 2000 $M_\Earth, 8$ to 32 $R_\Earth$) and 
    medium ($10$ to 100 $M_\Earth, 4$ to 8 $R_\Earth$) planets.}
%{Migrated evolution compared to data for $0.5 M_J$ and $0.1 M_J$}
   \label{fig:distrijupnep}
\end{center}
\end{figure}

\section{Discrepancy in $Q^{\prime}_{\ast}$ from giant and medium radii planets}
When tidal migration equations 
(\cite[e.g., Jackson et al. 2009]{jac09})
operate on an occurrence distribution 
of planets initially taken to be a single power law,
the result produced is a two-power law distribution, with 
tidal migration producing 
a short period fall off that follows a steeper power law.
In Fig.\,\ref{fig:distrijupnep},
we compare fits by H12 %\cite[Howard et al. (2012)]{how12} 
with fall offs calculated for a range of tidal distributions.
H12 used a two-power-law equation
to the fit Kepler planet candidates.
%We compare their fit with distributions for a range of tidal dissipation
%strengths $Q^{\prime}_{\ast}$, presented in Fig.\,\ref{fig:distrijupnep}
%in panels for large and medium radii planets.
We show their occurrence distributions for %of H12 for 
giant and medium radii planets
%two representative planet masses 
in Fig.\,\ref{fig:distrijupnep}, plotted against our calculations of 
occurrence distributions summed for mass
after a representive migration time of 4.5 Gyr,
shown for several values of tidal dissipation
strengths $Q^{\prime}_{\ast}$.
%at the representative age of 4.5 Gyr.
%and against these show the H12 
%\cite[Howard et al. (2012)]{how12} fit.  
The fall off for giant and medium planets could 
be interpreted as giving different values of
$Q^{\prime}_{\ast}$ for differently sized planets. 
The difference appears to correspond with the radii
where the pile up of 
giant planets occurs, which suggests that the flow into the pile up
raises the occurrence of the shortest period giant planets.
%NIL: We emphasize the relative difference in the location of the slope
%NIL: for large and medium planets, because the location of the curves 
%NIL: will move towards shorter or longer periods
%NIL: depending on the age distribution of 
%NIL: the stars, but we assume the age distribution of giant
%NIL: and medium planets to be the same.
%NIL: and the relative distribution will vary a little with
%NIL: the actual mass distribution, 
This difference appears to be best reconciled
by a flow of high eccentricity giant planets creating the pile up
%(\cite[Socrates et al. 2012]{soc12}), 
(S12), from which an increased number
of planets migrate though the shortest period region and into the star.

%(Fig.\,\ref{fig1}).

\begin{figure}[htb]
% \vspace*{-2.0 cm}
\begin{center}
%\includegraphics[width=\columnwidth]{fallinrate_3ranges_future_iauBJ.eps}
%Path.eps}
\includegraphics[width=\columnwidth]{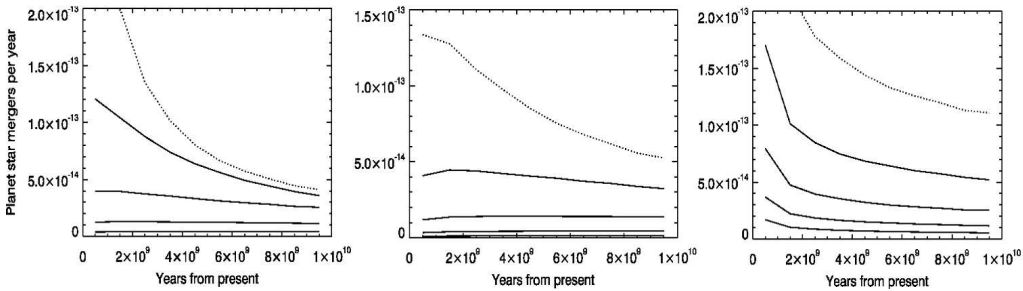}%1350-1836.eps}
% \includegraphics[width=4.0in]{fallinrate_3ranges_future_iauBJ.eps}%Path.eps} 
% \vspace*{-1.0 cm}
 \caption{Future fall in rates for the three ranges of planet radii 
  (three panels, for large, medium, small) each with plots for tidal
  dissipation values of $\log(Q^{\prime}_{\ast})$ 
  values of $10^{6.5}$ (dotted, top), 
  $10^{7.0}$, $10^{7.5}$, $10^{8.0}$, and $10^{8.5}$.}
   \label{fig:fallinrate}
\end{center}
\end{figure}

\section{Future fall in rates consistent only with giants arriving}

In the three panels of Fig.\,\ref{fig:fallinrate}
we show, for giant, medium, and relatively small planets,
the calculated future in fall rates plotted for 
% range of 
stellar tidal dissipation values of ${Q^{\prime}_{\ast}}$
from $10^{6.5}$ to $10^{8.5}$. 
%The belief has been that the distribution of these planets started with
%a shape similar to the occurence distribution that we see today, 
We show here the rate of infall calculated for the fits by H12;
%\cite[Howard et al. (2012)]{how12}; 
infall calculated directly using data directly,
though noisy,
show the same discrepancy (Taylor 2012, in preparation).
The value of $Q^{\prime}_{\ast}$ chosen must lead to infall rates
that decrease at a rate no faster than 
%NIL2: it can be believed that 
the supply of planets can reasonably decrease as the star population becomes older. 
%NIL2: It can be assumed that at any time, 
%NIL2: the rate should be constant for a constant age distribution.
%NIL2: For tidal dissipation $Q^{\prime}_{\ast}$ 
%NIL2: stronger than below $10^7$, it can be seen that for giant planets 
%NIL2: that extrapolating this curve back in time 
%NIL2: would require an unphysically large planet population.
Adding an inward flow of giant planets
%NIL2: can reduce the decrease such that an 
allows an extrapolation of the future
rate of infall such that $Q^{\prime}_{\ast} < 10^7$ 
is reasonable, which would be more consistent with
%NIL2: would not be too much less than today.
%NIL: The curve could become flat enough that extrapolating it 
%NIL: backwards would not require too much larger of a past population.
%NIL2: We find adjustment necessary because 
the tidal dissipation strength
evaluated for medium planets of closer to
$Q^{\prime}_{\ast} \sim 10^7$.
The rate of this flow  only needs to be on the
order of $10^{-12}$ giant planets per year per star
for the fall in rates to be consistent, a rate that the population
of longer period planets should be able to supply.
%NIL: The supply of outer planets should be able to sustain this rate for the life of 
%NIL: the stars mostly under consideration.

%NIL: We also have used parameters for the 
%NIL: Kepler planet candidates to calculate
%NIL: the fall in rates. The results are noisy, 
%NIL: but confirm the result here that 
%NIL: a small rate of flow is all that is 
%NIL: necessary to signficantly change
%NIL: the calculated strength of tidal dissipation.

\begin{figure}[tb]
%\begin{SCfigure}
%\centerline{\includegraphics[width=2.4in]%[width=3.4in]%[width=\columnwidth]
%2.4 in
\centerline{
  \includegraphics[width=2.3in]%[width=3.4in]%[width=\columnwidth]
    {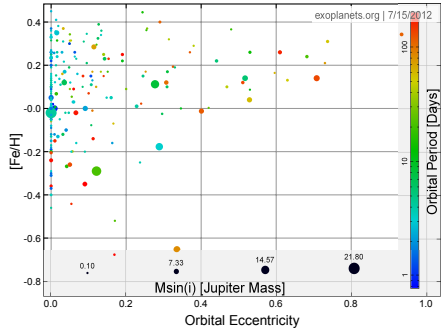}
}
\caption{
Correlation of increasing Fe/H with increasing eccentricity for 
planets with periods less than 200 days and masses greater than 0.1 $M_J$. 
Found using exoplanets.org. 
}
\label{fig:FehCorrEcc}  %Fe/H Correlation with Eccentricity
\end{figure}
%\end{SCfigure}
%From C:\Users\Stuart\Dropbox\talks_db\2012planetDistributionPapers\2012exoplanetsOrgOUT\desertMetalVsEccLT200dayGT0p1Mj.eps  
%ThinkPad T43p (2687-D5U) ATI Mobility FireGL V3200 ati2dvag.dll

\section{Multiplanet pollution, and conclusions}

If inwardly migrating giant planets are more likely to 
scatter other planets into the star
before circularizing, then this might explain why
short period giant radii planets are more often the only planet,
yet smaller planets are more often in multiple systems
(\cite[Fabrycky et al. 2012]{fab12}).
We present an apparent
correlation of higher eccentricity with increasing [Fe/H],
as shown in 
Fig.\,\ref{fig:FehCorrEcc}.
%NIL2: for planets of mass $>0.1 M_J$ and periods $<200$ days.
\cite[Buchhave  et al. (2012)]{buc12} find that stars hosting
smaller radii planets do not preferentially have higher [Fe/H].
Might the migration of large planets have disrupted orbits of smaller planets?
The pollution explanation for the correlation of [Fe/H] with planets
has been opposed based on missing correlations with other
elements, but it may be that circularized planets
have had more time for the star to have mixed the pollution.
%NIL2: than planets still in high eccentricity orbits.

%repeat Short period giant planets have been found to generally be the only planets
%repeat found around their host stars, while smaller planets are more often found in
%repeat multiple systems (\cite[Fabrycky et al. 2012]{fab12}).
%NIL2 such that they became pollution in the star?
%repeat Stars hosting giant planets have excess Fe/H,
%repeat while stars hosting planets of ``super-earth'' radii
%repeat (2 to 4 $R_\Earth$) do not (\cite[Buchhave  et al. 2012]{buc12}).
%repeat The mixed indicators of pollution in the shortest
%repeat period planets may be due to %there having been 
%repeat more time since the pollution event.
%NIL2: the shortest period orbits being at a
%NIL2: later stage of migration. 
%NIL3 (obvious):  We urge observations of stars with high eccentricity
%NIL3 (obvious):  orbits for signs of pollution.

The migration of  giant planets into short period orbits may have
disrupted the orbits of smaller planets.
It is important to understand why the flow of giant planets 
appears to be higher, %, and why there is a pileup of giant planets, 
but perhaps planets whose orbits have been disrupted migrate more quickly.

We conclude that there appears to be a flow of giant planets into the star
that appears to be significantly different from how smaller planets migrate
into the star.

%\bibitem[Anders \& Zinner (1993)]{AndersZinner93}
%{Anders, E., \& Zinner, E.} 1993, \textit{Meteoritics}, 28, 490
%\bibitem[Ott (1993)]{Ott93} {Ott, U.} 1993, \textit{Nature}, 364, 25
%\bibitem[Zinner (2004)]{Zinner04} 
%{Zinner, E.} 2004, in: K.K. Turekian, H.D. Holland \& A.M. Davis (eds.), 
% \textit{Treatise in Geochemistry 1} (Oxford and San Diego: Elsevier), p.\,17

\end{document}